\documentclass{openjournal}
\usepackage{amssymb,amsmath}
\def\app#1#2{%
	\mathrel{%
		\setbox0=\hbox{$#1\approx$}%
		\setbox2=\hbox{%
			\rlap{\hbox{$#1\propto$}}%
			\lower1.1\ht0\box0%
			}%
			\raise0.25\ht2\box2%
			}%
			}

\usepackage{revsymb,graphicx,amsmath,amssymb}
\begin{document}
\title{Transient X-ray Sources as Extremely Eccentric Mass-Transfer Binaries
with Compact Companions}
\shorttitle{Transient X-ray Sources as Extremely Eccentric Binaries}
\shortauthors{Katz \& Nowak}
\author{J. I. Katz}
\affil{Department of Physics and McDonnell Center for the Space Sciences,
Washington University, St. Louis, Mo. 63130}
\email{katz@wuphys.wustl.edu}
\author{M. A. Nowak}
\affil{Department of Physics and McDonnell Center for the Space Sciences,
Washington University, St. Louis, Mo. 63130}
\begin{abstract}
It has long been suggested that X-ray transients are produced at periastron
of stellar-compact object binaries with eccentric orbits.  Recoil of matter
evaporated from the star by X-rays from matter transferred at periastron
increases the orbital semi-major axis and eccentricity.  After periastrons
the object would be a transient X-ray source, a Galactic analogue of a tidal
disruption event (TDE), but recurrent with a gradually increasing period
rather than catastrophic.
\end{abstract}
\keywords{Black holes; neutron stars; binary orbits; transient X-ray
sources}
\section{Introduction}
Many authors \citep{CP75,AFP76,B76,OS77,M77,R78,AG80} have suggested that
transient X-ray sources may be the result of episodic mass transfer to
compact objects around periastrons of eccentric binaries.  Supergiant X-ray
Transients A0538-66 \citep{D83,BW86,S88,RL98}, XTE J1829-098 \citep{HG07},
SAX J1818.6-1703 \citep{ZC09}, IGR J18483-0311 \citep{R10}, XTE J1739-302
\citep{DCB10}, IGR J17544-2619 \citep{N13}, IGR J00370+6122 \citep{GNC14}
and IGR J11215-5952 \citep{LNC14} are known to have moderate eccentricities,
and might evolve to or be related to extremely eccentric systems.

\citet{N26} recently claimed, using the Doppler shift of the very stable
pulsation period of the $\delta$-Scuti star as a clock, that BE Lyncis (HD
79889) is in a binary orbit of period $P \approx 15.9\,$y and eccentricity
$e = 0.9989^{+0.0008}_{-0.0021}$ about a black hole of $M \ge 17.5\,M_\odot$.
This result was disputed by \citet{Na26}, but raised the question of if and
how orbits with such extreme eccentricity might be produced.

Here we consider a possible mechanism of producing extreme eccentricities in
mass-transfer binaries containing a compact object, independent of the
reality of the claims of \citet{N26}.  The mechanism more generally leads to
increasing eccentricity and semi-major axis to extreme values in binaries
with suitable initial conditions, ultimately unbinding them.

We first argue that it is very unlikely that a single impulsive kick would
produce extreme eccentricity from a more circular orbit.  That would require
precise tuning of the magnitude of the kick to leave the system bound, but
so very weakly bound.  We then suggest a mechanism to gradually increase the
eccentricity with each periastron passage: Mass transfer to the compact
object makes an accretional X-ray source.  The X-rays impinging on the
companion star drive a wind from its surface, and the star recoils.
Launched in the direction of the black hole from the X-ray illuminated
hemisphere, it does not change the orbital angular momentum but adds energy,
increasing the semi-major axis and eccentricity.  An essential feature of
this hypothesis is that the X-ray emission and recoil occur after
periastron, and not, {or much less,} before.

Such a system may remain a weak X-ray source for an extended time as the
disc left by the last mass transfer episode dissipates.  It is a strong
transient X-ray source after each periastron passage.  In the case of BE
Lyn the star is pulsating and the quantity of mass transferred and the
strength of the X-ray source would depend on the pulsational phase at
periastron, but most such hypothetical systems would have non-pulsating
stars.
\section{Single Kick?}
We do not have a theory for the possible distribution of kicks resulting
from the formation of the black hole, so can only make rough estimates.
The impulse per unit mass given to the companion star of mass $m$ required
to disrupt a binary of masses $M$, $m \ll M$ and initial semi-major axis
$a_0$ is $p_d \approx \sqrt{GM/a_0}$.  In order to leave it in an orbit of
semi-major axis $a = a_0/(1-e) \gg a_0$ with residual binding energy
$GM/2a$ the impulse per unit mass must be tuned to a range of width
$\Delta p \approx \sqrt{GM(1-e)/a_0}$.  The fraction of impulses, assumed
broadly distributed over a range of width $\sim p_d$, that fall within that
tuning range is ${\cal O}[(1-e)^{3/2}] \sim 2 \times 10^{-4}$ (for the
claimed parameters of BE Lyn).  This is small enough that we should look for
alternative mechanisms.
\section{Cumulative Periastron Passages}
At periastron, mass may be transferred from the star to the vicinity of the
compact object, where it forms an X-ray emitting accretion disc.  We do not
know how much mass is transferred (in BE Lyn it would depend on the
pulsational phase at periastron).  As a feasibility check of the hypothesis
we assume that the power radiated by the accretional flow equals the
Eddington luminosity $L_E$ of the compact object, using bounds found by
\citet{N26} as the values for a hypothetical system.

When the star is at a distance $r$ from the X-ray source of mass $M_X$ in
its nearly parabolic (near the compact object) orbit its speed
\begin{equation}
v = \sqrt{2GM_X \over r}.
\end{equation}
We can rewrite this as
\begin{equation}
dt = \sqrt{r \over 2GM_X}dr.
\end{equation}
The recoil force of matter heated by the X-rays and escaping the star,
assuming all the incident X-ray energy is converted to lifting this matter,
\begin{equation}
\label{F}
F = {L \pi R_s^2 \over 4 \pi r^2}{1 \over v_{recoil}},
\end{equation}
where $L$ is the accretional luminosity of the compact object and the recoil
velocity (neglecting any kinetic energy the matter has after leaving the
star's gravitational field)
\begin{equation}
v_{recoil} = \sqrt{2 G M_s \over R_s},
\end{equation}
where the subscript $s$ refers to the star.  The radiative ablation force
(Eq.~\ref{F}) does work on the orbit.  In a very eccentric orbit the star's
velocity is almost parallel to its direction to the black hole and to $F$,
so the increment to its orbital energy
\begin{equation}
\label{DE}
\Delta E \approx \int_{r_0}^\infty\!F\,dr \approx \int_{r_0}^\infty\!
{L R_s^2 \over 4 r^2 v_{recoil}}\ dr = {L \over 4}
{R_s^{5/2} \over \sqrt{2GM_s}} \int_{r_0}^\infty\!{dr \over r^2} =
{L \over 4}{R_s^{5/2} \sqrt{2GM_s}} {1 \over r_0},
\end{equation}
where $r_0$ is the separation at periastron.

In BE Lyn the parameters of \citet{N26} imply $\Delta E/M_s \approx 1.1
\times 10^8$ cm$^2$/s$^2$.  This may be compared to its present $E/M_s
\approx - 5 \times 10^{12}$ cm$^2$/s$^2$, suggesting an orbital evolution
time scale of $\sim 5 \times 10^4$ orbits, or $\sim 10^6$ y, after which
rapid expansion would lead to dissolution of the binary.  Although these
values are based on disputed orbital parameters, they indicate plausible
orders of magnitude.  Analogous estimates may be made for any extremely
eccentric system.  However, our assumption of Eddington-limiting X-ray
luminosity and that irradiation begins at a separation $r_0$ are uncertain,
and much lower $L$ and $\Delta E$ are possible.

The energy increment occurs once per orbit, with a period
\begin{equation}
T \approx 2 \pi \sqrt{a^3 \over GM_X},
\end{equation}
where we make the approximation $M_X+M_s \approx M_X$.
The (mean over many orbits) rate of change of orbital energy
\begin{equation}
{dE \over dt} \approx {\Delta E \over T}.
\end{equation}
Using $a = -GM_X/(2E)$ (again approximating $M_X+M_s \approx M_X$)
\begin{equation}
{da \over dt} \approx {GM_X \over E^2}{dE \over dt},
\end{equation}
where $t \gg T$ is the time since periastron became close enough for mass
transfer.  The ablation force is directed from the compact object and center
of mass, so that angular momentum is conserved.  The star's speed at
periastron changes little from orbit to orbit as its weakly bound orbit
becomes even more weakly bound and $r_0$ also remains nearly constant.  Then
\begin{equation}
\label{r0a}
r_0 = a(1-e)
\end{equation}
implies the evolution of eccentricity
\begin{equation}
1-e \approx {r_0 \over a}.
\end{equation}

These results imply, as $a \to \infty$ and $e \to 1$, the scaling laws
\begin{equation}
\label{summary}
\begin{split}
{dE \over dt} &\propto a^{-3/2}\\
{da \over dt} &\propto a^{1/2}\\
a &\propto t^2\\
1-e &\propto t^{-2}\\
P &\propto t^3,\\
\end{split}
\end{equation}
where $P$ is the orbital period.  This evolution will continue until the
orbit becomes so weakly bound that it is disrupted by encounters with
passing stars.  Until that happens, such systems would resemble common
proper motion binaries if the compact objects and their discs are
observable, but the latter may be too faint to observe in visible light.
The discs would be luminous transient X-ray sources following periastron,
and possibly weak persistent X-ray sources as the accretion disc gradually
dissipates.
%
\section{Consistency}
Here we verify that the mass loss to the compact object required to
grow the orbit and unbind the binary is only a small fraction of the stellar
mass $M_s$.  The orbital binding energy at initial semimajor axis $a_i$ and
eccentricity $e_i$ is
\begin{equation}
|U| = {GM\mu \over 2 a_i} = {GM\mu \over 2 r_0}(1-e_i),
\end{equation}
where $M$ is the total mass and $\mu$ the reduced mass.  The accretional
(X-ray) energy absorbed by the star after transfer of mass $\Delta M$
\begin{equation}
\label{Eabs}
E_{abs} \lesssim \epsilon \Delta M c^2 \left({R_s^2 \over 4 r^2}\right)
\end{equation}
where we approximate the ablation as occurring at a distance $r$ from the
compact object and $\epsilon \approx 0.1$ is the fraction of accreted rest
mass converted to radiation by the compact object.

Requiring that
\begin{equation}
\Delta E \sim E_{abs} {v_{orb} \over v_{recoil}} \gtrsim |U|,
\end{equation}
yields
\begin{equation}
{\Delta M \over \mu} \gtrsim {GM \over 2r_0} {4 r^2 \over R_s} {1 \over
\epsilon c^2} {v_{recoil} \over v_{orb}} (1-e_i) \sim 2 \times 10^{-3}
(1-e_i),
\end{equation}
where we have taken $4r^2/R_s^2 \sim 100$, a ratio of velocities $\approx 1$
and $M/r_0$ typical of main-sequence stars.  This condition is not
restrictive; it is satisfied for any $e_i$

We also verify that transferring $\Delta M$ is sufficient to drive that
mass loss by ablation:
\begin{equation}
\label{DM}
\Delta M \sim {2 E_{abs} \over v_{recoil}^2}.
\end{equation}
Substituting Eq.~\ref{Eabs} into Eq.~\ref{DM} and cancelling $\Delta M$
yields
\begin{equation}
\label{consistency}
1 \lesssim 2 \epsilon {c^2 \over v_{recoil}^2} {R_s^2 \over 4 r^2}.
\end{equation}
For plausible values the right hand side of Eq.~\ref{consistency} is
$\sim 10^2\text{--}10^3$, verifying consistency.

This consistency check might seem to imply that every binary with a compact
object should have runaway eccentricity and soon become unbound.  One reason
this is not so is that the proposed mechanism only works if the orbit is
initially eccentric enough that ablation post-astron exceeds ablation
pre-astron; that the accretion disc's luminosity fades before the star
returns to its next periastron.  This is difficult to quantify because we do
not know how fast that fading occurs, but {a rough estimate is possible.
The disc dissipation time is roughly
\begin{equation}
t_{disc} \sim {r_d \over \alpha h_d}\sqrt{r_d^3 \over GM_X},
\end{equation}
where $h_d$ is the disc's thickness, $r_d$ its radius and $\alpha$ the ratio
of viscous stress to pressure in the disc.  Conventional estimates are that
$r_d/(\alpha h_d) \sim 100$.

This should be compared to the orbital time $t_{orb} = \sqrt{a^3/GM_X}$; the
disc and X-ray emission dissipate before the companion star returns if
\begin{equation}
\label{compare}
t_{disc} \lesssim t_{orb}.
\end{equation}
The ratio $r_d/r_0 \sim 0.1$, depending somewhat but not sensitively on the
mass ratio \citep{K73,K24}.  Then Eqs.~\ref{r0a} and \ref{compare} imply
\begin{equation}
\label{emin}
e \gtrsim 0.5
\end{equation}
is required for the proposed mechanism to be effective.}

Another reason {that only initially eccentric compact object binaries
are disrupted by this mechanism} is that the increase of eccentricity by
ablation must exceed its decrease by tidal dissipation.  Dissipation may be
rapid in convective regions as a result of turbulent viscosity, but this is
also difficult to quantify.  When tidal distortion only occurs briefly
around periastron, likely on a shorter time scale than convective overturn,
the stellar deformation may be nearly adiabatic, with little dissipation.
{In radiative regions (where matter is stably stratified) dissipation is
very small; most of the volume of early-type supergiants is radiative and
convection only occurs in their cores where tidal distortion is small.}
\section{Discussion}
We propose that some Galactic transient X-ray sources are extremely
eccentric binaries containing a nondegenerate star and a compact object
(neutron star or black hole).  The {\it apparent} rarity of very eccentric
orbits in binaries containing compact objects may be attributed to the
difficulty of distinguishing them from single stars because of the faintness
of the compact objects, {to the fact that they spend most of their time
at large separations where their accelerations and orbital velocities are
small}, and to their limited lifetime as bound systems as their semimajor
axes rapidly increase (Eq.~\ref{summary}) before disruption by passing
stars.

The mechanism proposed here requires that the recoil during the outward
(post-periastron) part of the orbit must exceed any recoil during the inward
(pre-periastron) part of the orbit sufficiently to overcome any
circularizing process, such as tidal dissipation around periastron.  This is
difficult to quantify because the time-dependent dynamics of accretion discs
is not understood quantitatively, but qualitatively requires that the
accretion disc time scale be only a fraction of the orbital period.  This
implies that the initial eccentricity must be substantial in order that
radiation ablation drives its growth.  That is expected to be unusual
because mass-transfer binaries, and binaries close enough to evolve to
mass-transfer systems, tend to circularize by tidal dissipation.

The proposed mechanism assumes a periastron distance $r_0$ close enough
to the stellar radius that mass transfer is sufficient to produce a luminous
X-ray source by accretion onto the compact object, but not so close that
mass transfer is catastrophic.  This does not require tuned parameters
because main-sequence stars swell as they age.  Provided the initial orbital
eccentricity is large enough, unless the initial periastron distance exceeds
the maximum stellar radius on its evolutionary track the process described
here must begin at some time during its evolution and the eccentricity and
semi-major axis will grow.

{The mass transfer events at periastron proposed here may resemble the
micro-TDE proposed by \citet{R25}.  They differ in that in an eccentric
binary whose nondegenerate star is slowly evolving to a radius at which mass
transfer occurs at periastron, the mass transferred is only enough to drive
the orbit to larger semi-major axis, while in micro-TDE the periastron
distance is likely to be well within the star, and mass transfer is much
larger or even disruptive.}

Eq.~\ref{emin} suggests that compact object-nondegenerate star binaries with
$e \gtrapprox 0.8$ and periastrons close enough for mass transfer (and hence
X-ray sources) are rapidly disrupted.  Such systems should be rare, a
prediction that may be compared to the distribution of observed
eccentricities. 

In the model proposed here extremely eccentric mass-transfer binaries with
compact companions are luminous transient X-ray sources following their
periastrons.  If the orbital period is long, only one periastron may have
occurred within the historical X-ray database.  The rapid increase of $P$
(Eq.~\ref{summary}) is a testable prediction, provided that $P$ is not too
long, but implies that transient outbursts are not periodic.  The duration
and delay of emission after periastron depend on poorly understood accretion
disc dynamics, but may be at least an order of magnitude longer than
$\sqrt{r_0^3/GM_X}$.  Study of this emission would throw light on accretion
disc dynamics.

The lingering remnants of an accretion disc produced at periastron might
emit detectable persistent X-ray emission.  Apparently single stars
associated with low luminosity X-ray sources, as well as X-ray transients,
are candidates for extremely eccentric binary orbits.
\section*{Post-Script}
The discovery \citep{Y26} that PSR J0653$+$0443 is in a 79 day binary orbit
with the extraordinary eccentricity $e = 0.992387$ may perhaps be explained
by the mechanism proposed here.  After the first neutron star formed, the
system was a transient X-ray source, and radiation recoil of its companion 
increased the eccentricity so that $1-e$ became comparable to its present
value of 0.007613.  The second supernova gave the second neutron star a kick
$\lesssim 20$ km/s (the formation of compact objects with such small kicks
is inferred from the presence of neutron stars in globular clusters
\citep{K75}) that left the system bound.  The present binding energy of $2.5
\times 10^{13}$ ergs/g (taking a total mass of $2.8 M_\odot$ and
$\sin{i}=1$) indicates that a kick of 20 km/s would have had very little
effect on the orbital parameters.  The present distance of closest approach
$a(1-e) = 5.8 \times 10^{10}$ cm may be consistent with the evolution to
supernova of a repeatedly (at periastron passages) stripped stellar core.

\end{document}